# Supporting Therapeutic Relationships and Communication about Mental Health

**David Coyle**
Interaction and Graphics Group,
Dept. of Computer Science,
University of Bristol,
Bristol BS8 1UB, UK
david.coyle@bristol.ac.uk

**Gavin Doherty**
School of Computer Science
and Statistics,
Trinity College Dublin,
College Green,
Dublin 2, Ireland
gavin.doherty@cs.tcd.ie

## ACM Classification Keywords
H.5.m [Information Interfaces and Presentation]: Miscellaneous – interdisciplinary design, mental health

## Introduction and background
Effective communication and strong therapeutic relationships are critical to successful mental health interventions. For example, in 1957 Carl Rogers, a pioneer of person-centred therapy, proposed that an empowering relationship could, in and of itself, create *"the necessary and sufficient conditions"* for positive therapeutic outcomes [1]. Whilst modern psychological theories no longer favour an exclusive focus on relationships, positive relationships and the dynamics of client-therapist communication remain cornerstones of mental health intervention theories. A more recent meta-review concluded that across all interventions models, irrespective of the theoretical approach, the quality of the relationship between therapists and clients is the second leading determinant of successful clinical outcomes [2].

Over the past ten years we (David Coyle and Gavin Doherty) have designed and evaluated a wide range to systems that provide support for psychological (or talk-based) mental health interventions [3]. Here we briefly consider two recent examples. In each case our aim was to enhance communication and reshape clinical practice in a manner that empowers patients. gNats Island is a computer game that supports face-to-face interventions for adolescents [4]. MindBalance is an online treatment programme for adults experiencing difficulties with depression [5].

**Face-to-face communication: gNats Island**

Adolescents experiencing mental health difficulties often react confrontationally, or not at all, to direct conversations with a therapist. When therapists work with younger children, play therapy often provides an effective way of engaging, indirectly, in therapeutic processes. However adolescents can also react negatively to traditional play therapy if they feel they are being treated as children.

gNats Island is a desktop computer game that aims to address this imbalance. It implements key aspects of Cognitive Behavioural Therapy (CBT) and provides an age appropriate, face-to-face intervention for adolescents aged 10-15. The game provides an overall narrative in which players visit a tropical island and meet a team of wild life explorers. Characters introduce mental health concepts using spoken conversation, animations, videos and questions regarding the player's own situation. Concrete metaphors are used to explain abstract CBT concepts. For example negative automatic thoughts, a key concept in CBT, are presented as little creatures called gNats that can sting people, causing negative thinking. Through a series of in-game conversations players learn new strategies for identifying and challenging negative thoughts. Metaphors such as catching, trapping and swatting gNats are used to describe this process.

In sessions a therapist and adolescent sit together at a computer. Rather than talking face-to-face with the adolescent, the therapist acts as a partner in their exploration of the game world. As such gNats Island represents a substantial reshaping of the traditional therapeutic interaction. It was intended that conversations with game characters would provide a context for more detailed conversations between the adolescent and therapist. Further, it was predicted that the game would help to reduce the difficulties many adolescents experience with face-to-face interventions and assist in creating a client-centred, fun and experiential process.

Therapists who used gNats Island with adolescents were very positive about the way in which the game changed the dynamics of the therapeutic interaction. They highlighted both specific factors (e.g. eye contact) and the general role of the game as a mediating factor:

*"I thought it was really good from an eye contact point of view, he doesn't like making a lot of eye contact, so having the screen to focus in on was perfect."*

*"It was almost like a transitional object or an external kind of mediating factor, so that I suppose the sessions were less directed, less challenging ... so the child found it easier to engage through the medium of the game."*

Clinicians also felt the game had a beneficial impact on the client-therapist relationship:

*"The most valid point … is that it enhanced engagement and thus the therapeutic relationship."*

Overall clinicians valued the way that gNats Island supported and even enhanced two-way conversations between therapists and clients. Further details of our evaluation of gNats Island are available in [4].

**Online treatment and support: MindBalance**

While the efficacy of many face-to-face interventions is well demonstrated, limitations in the availability of therapists, coupled with the time intensive nature of treatments, effectively means that only a minority of people can receive the treatment and support they need. For working people and those in remote locations, the options become even more limited.

Online mental health treatments have the potential to help address these issues [3]. However, problems with adherence are often encountered in real-world deployments of standalone online interventions. Without the support of a therapist there is a high probability that people will not engage with their treatment or will simply drop out of their programme. A more promising strategy is to incorporating low-intensity support and communication into online treatment programmes. MindBalance is a CBT-based programme for depression, which takes this approach [5]. It is delivered using the SilverCloud e-health platform. Many of the core features of the platform relate to mediation of the interaction between clients and clinicians. For example, clients are encouraged to share information entered into the system with their supporter, often a clinician or mental health assistant. This can include a personal profile and the details of therapeutic exercises (a thoughts-feelings-behaviours chart for example). The supporter then provides individual feedback.

Qualitative feedback from the evaluation of MindBalance [5] indicates that, for many clients, providing a human presence helps engagement (their efforts were not going unnoticed) and gives a sense of support, even if they do not interact with their supporter to any significant extent. It does however require the supporter to establish some degree of social presence. One open and interesting research question relates to the amount of personal information supporters should share about themselves. From a HCI perspective, we might favour some level of reciprocity (providing a photo and details of qualifications for example). However others have advanced the hypothesis, following studies on lean communication media, that providing less information will lead clients to form more positive, idealised views of communication partners [7].

While MindBalance was initially envisioned as an online intervention, therapists with whom we work also chose to use it as an adjunct to face-to-face treatment. Perhaps surprisingly, preliminary data shows a high degree of engagement with the online system in such cases. With adjunct use, alongside seeing their therapist face-to-face, clients use the online system between sessions. Online activities supplement and provide material for face-to-face meetings. Initial results suggest this approach can play a number of roles in facilitating communication. For example, in our previous research on mobile mood diaries, one specific use of the technology was to broach sensitive subjects, not easily raised in person [6]. The nature of the communication in MindBalance may also facilitate this kind of disclosure, with one client commenting:

*"I also felt like I was unloading some of my problems to another person without having to say it out"*.

A broader point relates to the issue of unloading problems. In MindBalance, being able to communicate asynchronously and record information online, and in the moment, rather than waiting for the next session was significant, even though feedback would not be received until the next scheduled review or face-to-face session.

## Conclusions

MindBalance and gNats Island have both now been widely used. In the case of MindBalance this includes use as a low intensity online intervention and as an adjunct to face-to-face treatment. gNats Island has focused solely on supporting face-to-face treatments. Since the initial evaluations, reported at ACM CHI 2011 [4], over 750 mental health professionals in Ireland, the UK and the US have received training in the gNats Island interventions for adolescents.

Ultimately we believe mental health services will benefit from an integrated approach, which utilises a broad range of technologies, to support online and face-to-face communication and treatment. One specific aim for such an approach is to tailor the intensity of face-to-face treatment based on the severity of difficulties and the preferences of different clients.

## Citations

[1] Rogers, C., *The necessary and sufficient conditions of therapeutic personality change.* Journal of consulting and clinical psychology, 1957. **21**: 95-103.

[2] Assay, T.P. and M.J. Lambert, *The Empirical Case for Common Factors in Therapy: Quantitative Findings*, in *The Heart and Soul of Change*, B.L. Duncan, M.L. Hubble, and S.D. Miller, Editors. 1999, American Psychological Association: 23-55.

[3] Coyle, D., et al., *Computers in Talk-Based Mental Health Interventions.* Interacting with Computers, 2007. **19**: 545-62.

[4] Coyle, D., et al., *Exploratory Evaluations of a Computer Game Supporting Cognitive Behavioural Therapy for Adolescents*, *ACM CHI* 2011, 2937-2946.

[5] Doherty, G., el al., *Engagement with Online Mental Health Interventions: An Exploratory Clinical Study of a Treatment for Depression*, *ACM CHI* 2012, 1421-30.

[6] Matthews, M., & Doherty, G. In the mood: engaging teenagers in psychotherapy using mobile phones, ACM CHI 2011, 2947-56.

[7] Mohr DC, Cuijpers P, Lehman K Supportive Accountability: A Model for Providing Human Support to Enhance Adherence to eHealth Interventions. J Med Internet Res 2011, 13(1).